# Anisotropic character of the metal-to-metal transition in $Pr_4Ni_3O_{10}$

Shangxiong Huangfu[*], Gawryluk Dariusz Jakub[†], Xiaofu Zhang[*], Olivier Blacque[‡], Pascal Puphal[†], Ekaterina Pomjakushina[†], Fabian O. von Rohr[‡], Andreas Schilling[*]

[*]Department of Physics, University of Zurich, Winterthurerstrasse 190 CH-8057 Zurich Switzerland
[†]Laboratory for Multiscale Materials Experiments (LMX), Paul Scherrer Institute (PSI), Forschungsstrasse 111, CH-5232 Villigen, Switzerland
[‡]Department of Chemistry, University of Zurich, Winterthurerstrasse 190 CH-8057 Zurich Switzerland

**ABSTRACT**

As a member of the Ruddlesden-Popper $Ln_{n+1}Ni_nO_{3n+1}$ series rare-earth-nickelates, the $Pr_4Ni_3O_{10}$ consists of infinite quasi-two-dimensional perovskite-like Ni-O based layers. Although a metal-to-metal phase transition at $T_{pt} \approx 157$ K has been revealed by previous studies, a comprehensive study of physical properties associated with this transition has not yet been performed. We have grown single crystals of $Pr_4Ni_3O_{10}$ at high oxygen pressure, and report on the physical properties around that phase transition, such as heat-capacity, electric-transport and magnetization. We observe a distinctly anisotropic behavior between in-plane and out-of-plane properties: a metal-to-metal transition at $T_{pt}$ within the *a-b* plane, and a metal-to-insulator-like transition along the *c*-axis with decreasing temperature. Moreover, an anisotropic and anomalous negative magneto-resistance is observed at $T_{pt}$ that we attribute to a slight suppression of the first-order transition with magnetic field. The magnetic-susceptibility can be well described by a Curie-Weiss law, with different Curie-constants and Pauli-spin susceptibilities between the high-temperature and the low-temperature phases. The single crystal X-ray diffraction measurements show a shape variation of the different $NiO_6$ octahedra from the high-temperature phase to the low-temperature phase. This subtle change of the environment of the Ni sites is likely responsible for the different physical properties at high and low temperatures.

## I. INTRODUCTION

Quasi-two-dimensional (quasi-2D) transition-metal oxides generally show a variety of intriguing electronic properties due to their strongly correlated $d$-electrons [1, 2], electric anisotropy [3 - 6], as well as the mixed-valence states of the transition-metal ions [7, 8]. Such layered compounds can display high-temperature superconductivity [9, 10], colossal magneto-resistance [11, 12], charge stripes [13, 14], and other complex physical phenomena. There exist various transitions between different phases with distinct properties, e.g., temperature or pressure induced metal-to-insulator transitions [15 - 19], or a doping induced Mott-insulator to superconductor transition in the cuprates [9, 20, 21]. The Ruddlesden-Popper (R-P) rare-earth nickelates $Ln_{n+1}Ni_nO_{3n+1}$ ($Ln$ = rare-earth) have recently attracted a lot of interest because they contain quasi-2D Ni-O layers which are similar to the Cu-O layers in the high-temperature superconductors [22-25], and superconductivity has indeed been reported to occur in a structurally and electronically closely related compound $Nd_{0.8}Sr_{0.2}NiO_2$ with planar Ni-O layers [26]. The layers in $Ln_{n+1}Ni_nO_{3n+1}$ are piled up to an infinite perovskite-like layered structure. They are connected by $NiO_6$ octahedra via shared vertex (both equatorial and axial) oxygen ions, and separated by rock-salt like layers ($Ln$-O) along the crystallographic $c$-axis.

These rare earth nickelates have shown various phases at different temperatures and magnetic fields. To be specific, perovskite $Ln$NiO$_3$ ($Ln$ = Y, Pr to Gd, Dy to Lu, $n = \infty$) show a metal-to-insulator phase transition [16, 27, 28], which is accompanied by structural changes [29]. Close to the transition temperature, the pronounced electron-lattice interaction induces a distortion of the $NiO_6$ octahedra, leading to significant changes of the Ni-O bond lengths at different crystallographic Ni sites in the low-temperature phases. Ultimately, it is a charge disproportionation commensurate with the crystallographic unit cell that generates this phase transition [30-33]. Similarly, in the $Ln_2NiO_4$ series ($n = 1$), a first-order structural phase transition occurs in $La_2NiO_4$, $Pr_2NiO_4$, and $Nd_2NiO_4$, from a high-temperature orthorhombic phase to a low-temperature tetragonal phase at around 80 K [34], 115 K

[35], and 130 K [36, 37], respectively. By contrast, the only known compound of the $Ln_3Ni_2O_7$ ($n = 2$) series, $La_3Ni_2O_7$, shows a structural transition at 550 K [38]. However, it exhibits a metallic behavior with a weak "kink" around 128 K in the resistivity, and a downward "peak" in the magnetization near 128 K [39]. In the $Ln_4Ni_3O_{10}$ series ($Ln$ = La, Pr and Nd, with $n = 3$), analogous indications for transitions have been reported in terms of a resistivity shift in the temperature range of 140 - 165 K, and a slightly anomalous magnetization in the same temperature range in $La_4Ni_3O_{10}$ [24, 39 - 41]. Nonetheless, a comprehensive study of the physical properties near these transitions in the $Ln_4Ni_3O_{10}$ series is still lacking.

The significant in-plane and out-of-plane anisotropy in most R-P rare-earth nickelates leads to quasi-2D electronic and magnetic properties [42, 43]. Due to the high melting point of nickelates and their complex Ni valence states, the growth of corresponding single crystals is rather difficult [44], and generally necessitates the application of high oxygen pressure. This makes the study of the anisotropic properties of this compound series very difficult. While the successful single-crystal growth of $LaNiO_3$ has recently been reported, anisotropic properties are not expected in this compound with an isotropy in its crystal structure along the three pseudo-cubic directions [45, 46]. Single crystals of $Ln_2NiO_4$ ($Ln$ = La, Pr, Nd) have been synthesized by the skull-melting method [47], and the electric-transport measurements on $La_2NiO_4$ showed that the in-plane resistivity is three orders magnitudes smaller than the out-of-plane resistivity [48]. Single crystals of $La_4Ni_3O_{10}$ and $Pr_4Ni_3O_{10}$ have also been obtained recently [24, 49], and angle-resolved photoemission spectroscopy measurements on $La_4Ni_3O_{10}$ show a gap opening at a phase transition, which is likely associated with the occurrence of a charge-density wave (CDW), but no anisotropic properties have been reported so far [24].

In this report, we have successfully grown $Pr_4Ni_3O_{10}$ single crystals in a traveling solvent floating-zone furnace at high oxygen pressure. We have performed measurements on the crystal structure, the heat capacity, and the electric and magnetic properties, which all exhibit distinct changes at the phase transition temperature at $T_{pt} \approx 157$ K. We show that this transition is slightly field-dependent and of first order.

## II. METHODS

We firstly synthesized $Pr_4Ni_3O_{10}$ powder by a sol-gel method with citric acid assistance to obtain a pure phase at relatively moderate reaction conditions [50], and the precursor powder was annealed at 1100 ˚C in oxygen atmosphere for 24 hours. The single-crystal growth of $Pr_4Ni_3O_{10}$ was performed in an optical-image traveling solvent floating-zone furnace in an oxygen atmosphere at a pressure of $\approx 140$ bar. The obtained single crystals are as large as $3.3 \times 1.6 \times 0.3$ mm$^3$ (see the inset of Fig. 1c). The details concerning the synthesis procedures are described in the Supplemental Material [51]. The structural changes of the $Pr_4Ni_3O_{10}$ single crystals were characterized by single crystal X-ray diffraction (XRD) both at room temperature and at 150 K. An XRD measurement on pulverized single-crystal powder shows sharp peaks [Fig 1(c)] revealing the high sample quality from the reactions.

The electrical resistivity was measured with a Physical Property Measurement System (PPMS, *Quantum Design Inc.*) and a standard four-probe technique with electrical contacts attached with silver paste. The zero-field transport measurements were performed for both in-plane and out-of-plane configurations in the temperature range from 10 K to 300 K. The magneto-transport measurements were done in external magnetic fields up to $B = 9$ T, where $B$ was oriented perpendicular to the current. The heat-capacity was measured with the heat-capacity option of the PPMS in the temperature range between 10 K and 300 K in zero field and in $B = 7$ T. The magnetic properties were studied in a Magnetic Properties Measurement System (MPMS 3 from *Quantum Design Inc.*), which is equipped with a reciprocating sample option (RSO), with external fields $B$ both parallel and perpendicular to the *c*-axis.

# III. RESULTS

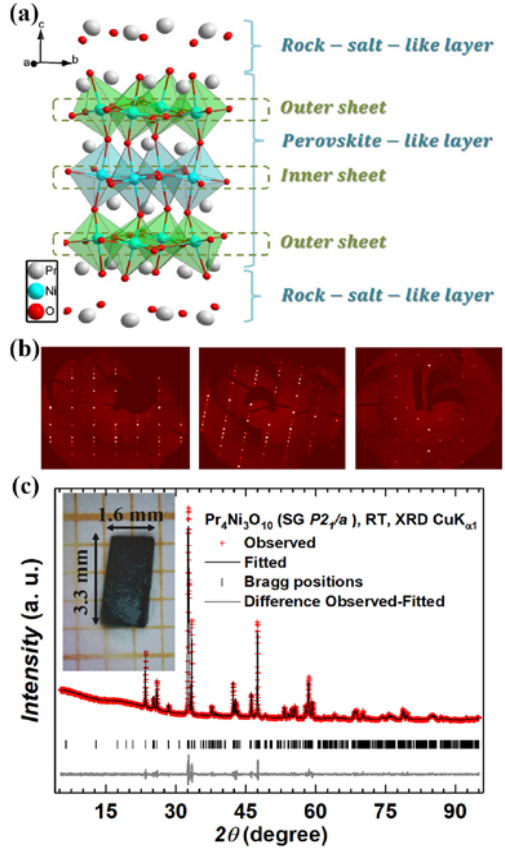

Fig. 1 (a) Illustration of the layered crystal structure of $Pr_4Ni_3O_{10}$; (b) ($0kl$), ($h0k$) and ($hk0$) planes (from left to right) of single crystal $Pr_4Ni_3O_{10}$ based on a single crystal diffraction experiment performed at 150(1) K ($P2_1/a$ settings); (c) Experimental powder XRD ($CuK_{\alpha 1}$ radiation) pattern (red crosses) for pulverized single crystal sample of $Pr_4Ni_3O_{10}$ at room temperature. The black line corresponds to the best fit from the Rietveld refinement analysis. Lower vertical marks denote the Bragg peak positions. The bottom, grey line represents the difference between experimental and calculated points; the inset of (c) shows the image of the crystal used for all the measurements.

In $Pr_4Ni_3O_{10}$, featuring $P2_1/a$ space group, there exist two different types of Ni sites in the crystal structure as it is illustrated in Fig. 1(a), representing the Ni sites in the inner perovskite layer ($Ni_{in}$) and in the outer perovskite layers ($Ni_{out}$) respectively. All the Ni ions are located in the center of the disordered $NiO_6$ octahedra, resulting in a monoclinic symmetry of the unit cell. It has been recently shown that upon lowering the temperature, the $b$-axis was expanded below $T_{pt}$, giving rise to a structural change at this transition temperature [49]. We therefore performed single-crystal XRD both at room temperature (above $T_{pt}$) and at 150 K (below $T_{pt}$), the detailed results of which are listed in Supplemental Material [51]. We verify a larger *b/c* value at 150 K than at room temperature, which is consistent with the previous results [49]. Furthermore, the bond lengths of Ni-O show distinctly different temperature dependences at the different Ni sites, which we summarized in Table I. To be specific, the $Ni_{in}$-O lengths clearly change between room temperature and 150 K, while there are no visible differences between the corresponding $Ni_{out}$-O bond lengths. We conclude that the inner $NiO_6$ octahedra shrink during the phase transition, while the outer $NiO_6$ octahedra remain almost unaffected.

Table I. Average Ni-O bond lengths at different Ni sites at different temperatures

|  | Room temperature ($T > T_{pt}$) | | 150 K ($T < T_{pt}$) | |
| --- | --- | --- | --- | --- |
|  | $<Ni-O_e>$ (Å)[I] | $<Ni-O_a>$ (Å)[I] | $<Ni-O_e>$ (Å) | $<Ni-O_a>$ (Å) |
| $Ni_{in}$ | 1.94 | 1.92 | 1.93 | 1.90 |
| $Ni_{out}$ | 1.93 | 2.08 | 1.93 | 2.08 |

[I] e = equatorial and a = axial

Fig. 2(a) shows the temperature dependence of the heat capacity *C* of the $Pr_4Ni_3O_{10}$ single crystal in zero magnetic field between 10 K and 300 K. The heat capacity reaches $\approx 390$ J mol$^{-1}$ K$^{-1}$ at 300 K, approaching the classical value from the Dulong-Petit model ($C = 3nR \approx 424$ J mol$^{-1}$ K$^{-1}$, where $R = 8.314$ J mol$^{-1}$ K$^{-1}$ is the gas constant and *n* is the number of atoms per formula unit). A clear *δ*-like peak appears at $T_{pt} \approx 157$ K, with a peak height in $C/T$ of $\approx 1.05$ J mol$^{-1}$ K$^{-2}$ and a peak width of $\approx 4$ K [see inset of Fig. 2(a)], which is considerably sharper than previously reported results [49]. By integrating $C/T$ vs. *T*, we obtain the entropy *S(T)* that displays an abrupt change ($\Delta S \approx 2.0$ J mol$^{-1}$ K$^{-1}$) at the phase transition [Fig. 2(b)], indicating a possible first-order phase transition. We note, however, that there is no detectable hysteresis around $T_{pt}$ within the temperature resolution of the measurement (≈ 0.5 K) [Fig. S1], and that a value of $\Delta S$ lower than $R\ln(2)$ is often observed at phase transition in metallic nickel oxides [52].

Considering the structural change at $T_{pt}$ [49] as it is revealed by our single crystal XRD measurements, we may expect to a variety of further changes of the physical properties of $Pr_4Ni_3O_{10}$ as we shall see below.

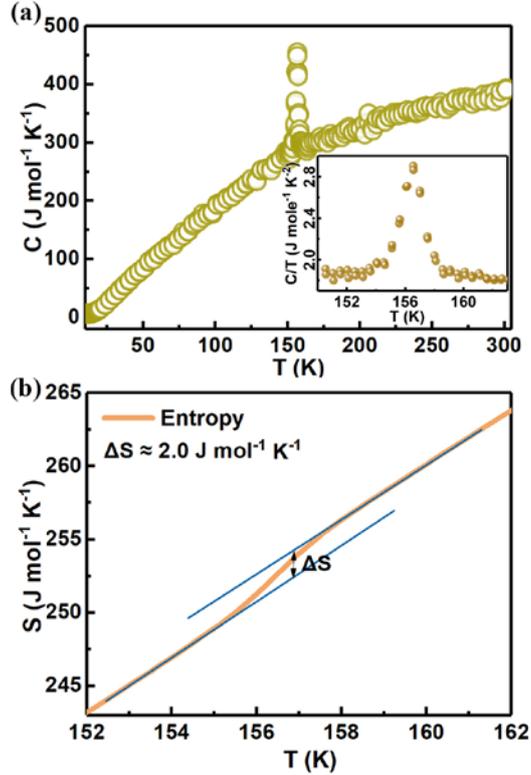

Fig. 2 (a) Temperature dependence of heat-capacity of $Pr_4Ni_3O_{10}$ between 10 K and 300 K. Inset: $C/T$ as a function of temperature around $T_{pt}$; (b) Corresponding entropy $S(T)$, showing a step-like feature at $T_{pt}$.

To clarify the influence of this phase transition on the electric-transport properties, we measured the temperature-dependent in-plane ($\rho_\parallel$) and out-of-plane ($\rho_\perp$) resistivities in zero magnetic field. Fig. 3 shows the temperature dependence of $\rho_\parallel$ and $\rho_\perp$ from 10 K to 300 K, normalized to the corresponding resistivities at 300 K. The absolute values at room temperature are roughly $\rho_\parallel \approx 5.6 \times 10^{-3}$ Ωcm and $\rho_\perp \approx$ 0.047 Ωcm, respectively, which is considerably less than reported for the oxygen-reduced single crystals $Pr_4Ni_3O_8$ (≈ 67 Ωcm) [25], but is in line with data published for polycrystalline $Pr_4Ni_3O_{10}$ (≈ 0.015 Ωcm) [40]. The ratio of the measured values at room temperature is $\rho_\perp/\rho_\parallel \approx 8.4$. In the high-temperature regime, these quantities show a similar metallic behavior. With decreasing temperature, however, both $\rho_\parallel(T)$ and $\rho_\perp(T)$ exhibit a sharp increase but a qualitatively different behavior below $T_{pt}$. Within the *a-b* plane, the resistivity $\rho_\parallel(T)$ shows a metal-to-metal transition, while along the *c*-axis, the $\rho_\perp(T)$ significantly increases with decreasing temperature over a broad temperature range, indicating rather a metal-to-insulator-like transition. We cannot rule out that the fact that $\rho_\perp(T)$ appears to decrease again in the low-temperature limit is due to a misalignment of the electrical leads and the irregular shape of the thin crystal, thereby mixing $\rho_\parallel$ and $\rho_\perp$ in the measured data to some extent. Therefore, the room-temperature ratio $\rho_\perp/\rho_\parallel \approx 8.4$ can only be taken as a lower bound of the resistance anisotropy.

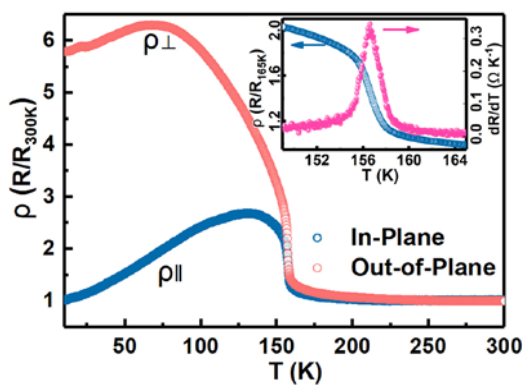

Fig. 3 Temperature dependence of the normalized zero-field resistivities of $Pr_4Ni_3O_{10}$, both in-plane ($\rho_\parallel$) and out-of-plane ($\rho_\perp$); the inset shows resistivity data $\rho_\parallel$ around $T_{pt}$ of both the normalized resistivity and its derivative, as functions of temperature.

Since the magneto-transport properties are closely related to the electronic structure near the Fermi surface, we also measured the magneto-resistance (MR) along different directions between 10 K and 300 K in detail. In Fig. 4(a) and (b) we show the resistance $R(B)$ for varying external magnetic field at constant temperatures $T$, normalized to the respective zero-field values $R(0)$. The response to the magnetic field is positive at low temperatures (with $\partial\rho/\partial B > 0$) for both in-plane and out-of-plane configurations, and becomes weaker with increasing temperature, which is very common for simple metals [53, 54]. In a very narrow region around the phase transition at $T_{pt}$, however, the MR effect becomes large and negative, i.e., with $\partial \rho / \partial B < 0$. In the Figs. 4(c) and (d) we are plotting the relative changes in resistance between zero field and $B$ = 9 T, indicating a seemingly anomalous sign change around $T_{pt}$. While outside this transition region, the normalized MR of the in-plane resistance $\rho_\parallel$ is larger than that of the out-of-plane resistance $\rho_\perp$, the negative MR around $T_{pt}$ is larger for perpendicular current transport.

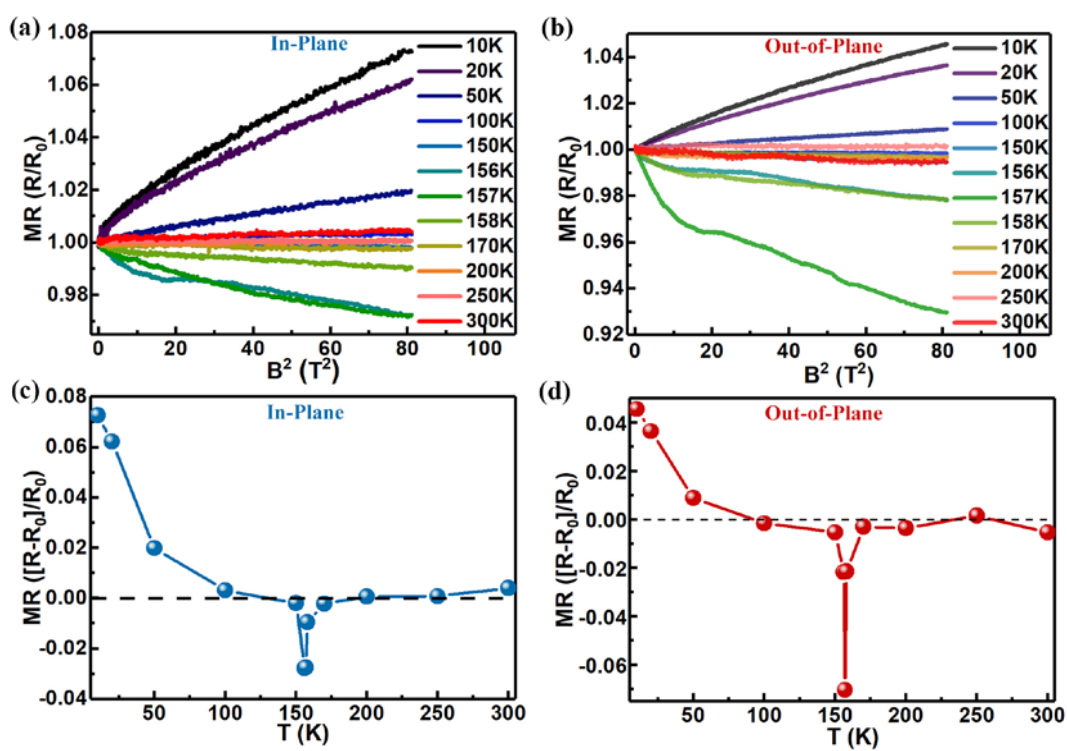

Fig. 4 The magneto-resistivity measurements of $Pr_4Ni_3O_{10}$ for both in-plane (a) and out-of-plane (b) configurations, plotted vs. $B^2$ for clarity; (c) and (d) show the respective temperature dependences of the magneto-resistivities for $B$ = 9 T.

The R-P $Ln_4Ni_3O_{10}$ nickelates have been reported to show a rather complicated temperature dependence of the magnetic behavior [42, 55]. In Fig. 5(a) and (b) we present the magnetic susceptibility near $T_{pt}$ for $B//c$ and $B\perp c$, showing a step-like feature at $T_{pt}$, which is less obvious when compared with the $La_4Ni_3O_{10}$ peer [39, 55], however. The temperature derivative of $M(T)$ allows for a clearer identification of the transition temperature [Fig. 5(a) and (b)]. The magnetic susceptibility for zero-field-cooling (ZFC) and field-cooling (FC) overlap nearly in the whole measurement range [Fig. S2], without any detectable hysteretic behavior. The temperature dependence of the magnetic susceptibility on both sides of the transition can be well fitted to a Néel-type law, $\chi(T) = \frac{C}{T+\Theta} + \chi_0$ [results are listed in Table II], demonstrating the paramagnetic nature of $Pr_4Ni_3O_{10}$ in the whole measured temperature range. As expected, the resulting Curie-constants $C$ on both sides of the transition are nearly field-independent and isotropic [see Table II and Fig. S3]. In the low-temperature phase, the Curie-constant is $C \approx 6.4$ emu K Oe$^{-1}$ mol$^{-1}$, which is compatible with the assumption that the magnetization in this temperature range is governed by the $Pr^{3+}$ ions with a magnetic moment of 3.58 $\mu_B$ per $Pr^{3+}$, i.e., the theoretical value for free $Pr^{3+}$ ions. Similar values have been reported for $PrNiO_3$ (3.31 $\mu_B$) [56] and $Pr_2NiO_4$ (3.73 $\mu_B$) [57]. The Ni ions do not seem to significantly contribute to the magnetic-susceptibility at low temperatures. In the high-temperature phase, the resulting Curie-constant ($C \approx 7.6$ emu K Oe$^{-1}$ mol$^{-1}$) is $\approx$ 18% larger, indicating the presence of an additional contribution from the Ni ions to the magnetization.

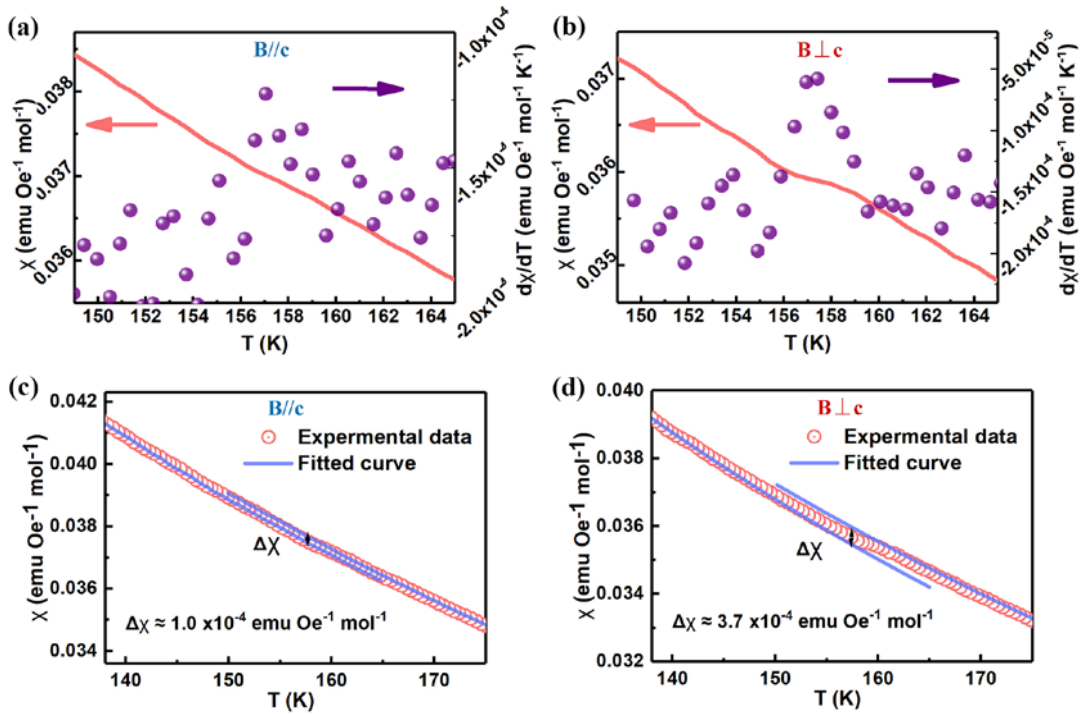

Fig. 5 Field-cooling (FC) magnetic-susceptibility of $Pr_4Ni_3O_{10}$ and its derivative in an external magnetic field $B$ = 0.1 T near the phase transition for $B//c$ (a) and $B \perp c$ (b); Magnetic-susceptibility data with Néel-type fits above and below $T_{pt}$ for $B//c$ (c) and $B \perp c$ (d); $\Delta\chi$ corresponds to the step-like increase of the magnetic susceptibility from the low-temperature to the high-temperature phase; the corresponding zero-field cooling (ZFC) data overlap with no detectable hysteresis and are therefore not shown for clarity.

Table II. Fitting results of the $\chi(T)$ data

|  |  | $C$ (emu K Oe$^{-1}$ mol$^{-1}$) | $\chi_0$ (emu Oe$^{-1}$ mol$^{-1}$) | $\Theta$ (K) |
|---|---|---|---|---|
| Low-temperature phase | $B//c$ | 6.4(5) | 3.9(9) × $10^{-3}$ | 37±6 |
|  | $B \perp c$ | 6.4(1) | 2.2(7) × $10^{-3}$ | 35±1 |
| High-temperature phase | $B//c$ | 7.6(4) | 9(3) × $10^{-4}$ | 53±5 |
|  | $B \perp c$ | 7.6(3) | 2.5(9) × $10^{-4}$ | 56±4 |

The fitted values for the Pauli-paramagnetic susceptibility are quite large in the low-temperature phase, with $\chi_0^L \approx 3.9\times 10^{-3}$ emu Oe$^{-1}$ mol$^{-1}$ for $B//c$ and $\chi_0^L \approx 2.2\times 10^{-3}$ emu Oe$^{-1}$ mol$^{-1}$ for $B \perp c$, while they are substantially smaller at high temperatures above $T_{pt}$, with $\chi_0^H \approx 9\times 10^{-4}$ emu Oe$^{-1}$ mol$^{-1}$ for $B//c$ and $\chi_0^H \approx 2.5\times 10^{-4}$ emu Oe$^{-1}$ mol$^{-1}$ for $B \perp c$, respectively.

## IV. DISCUSSION

Although not detectably hysteretic, the phase transition at $T_{pt} \approx 157$ K in $Pr_4Ni_3O_{10}$ is of first order, and must be slightly field-dependent. This conclusion is supported by our combined heat-capacity, magnetic-susceptibility and MR data as we shall see in the following. The simplest explanation for the observed sign change of the MR around $T_{pt}$ [Fig. 4] is the assumption that the phase transition is shifted towards lower temperature with increasing magnetic field, thereby allowing the compound, once held at a temperature $T$ slightly below $T_{pt}$ in zero magnetic field, to re-enter the more metallic high-temperature phase with increasing field. From our resistance data [see Fig. 3] and assuming a parallel shift of $R(T)$ by $\Delta T_{pt}$ with increasing $B$, we estimate $\Delta T_{pt} \approx -0.1$ K in $B = 7$ T. Unfortunately, the temperature resolution of our heat-capacity data taken in zero magnetic field and in $B = 7$ T ($\approx 0.5$ K) is not sufficient to resolve such a small effect of the magnetic field on $T_{pt}$ [see Fig. S1]. However, the magnetic-susceptibility exhibits an virtually field-independent step-like increase at $T_{pt}$ of the order of $\Delta\chi \approx 1.0 \times 10^{-4}$ emu Oe$^{-1}$ mol$^{-1}$ for $B//c$ and $\Delta\chi \approx 3.7 \times 10^{-4}$ emu Oe$^{-1}$ mol$^{-1}$ for $B \perp c$, respectively [see Fig. 5(a), 5(b) and S5], indicating that the first-order phase transition at $T_{pt}$ as it is revealed by the heat capacity measurements must follow a non-trivial phase transition line $B_{pt}(T)$. If we relate the measured changes in magnetization at $T_{pt}$, $\Delta M = B\Delta\chi$, to the (apparently field-independent) entropy change $\Delta S \approx 2.0$ J mol$^{-1}$ K$^{-1}$ via the Clausius-Clapeyron equation $-\Delta M \left(\frac{dB_{pt}}{dT}\right) = \Delta S$ that is valid at magnetic first-order transitions, we obtain $B_{pt}(T) \approx \sqrt{2\Delta S(T_{pt(B=0)} - T)/\Delta\chi}$, i.e., the phase-transition temperature must indeed decrease with increasing magnetic field $B$. This equation, together with the measured values for $\Delta\chi$ and $\Delta S$, predicts a decrease of $T_{pt}$ by $\Delta T_{pt} \approx -0.045$ K in $B = 7$ T for $B \perp c$, which is in fair agreement with the above estimate from our MR data. It is therefore safe to conclude that the phase transition at $T_{pt}$, with its associated changes of the crystal structure and the physical properties, is field-dependent and must be of first order, in agreement with Ref. [40].

We suggest that the very similar magneto-resistance effect observed at the spin-density-wave/charge-density-wave instability in Na$_2$Ti$_2$Sb$_2$O around $T = 110$ K [58], can be explained in a similar fashion as being a consequence of a certain field dependence of this transition. We assume that such an effect is potentially also present in other systems showing field-dependent phase transitions that are accompanied by large resistance changes.

While the metallicity is preserved for in-plane transport at all temperatures, the out-of-plane resistivity is enhanced below $T_{pt}$ [Fig. 3], which indicates either a metal-to-insulator-like transition for the respective crystal direction, or at least a significant increase of the anisotropy between in-plane and out-of-plane properties induced by the structural phase transition at $T_{pt}$. With an average valance state of +2.67, it is very likely that certain $E_g$ orbitals of the Ni ions are crossing the Fermi level, thereby forming partially filled energy bands. The situation is somewhat complicated by the fact that the inner $NiO_6$ octahedra are contracted in the *c*-direction, while the outer ones are elongated [see Table I], leading to a reversed hierarchy of the respective $d_{x^2-y^2}$ and $d_{3z^2-r^2}$ orbitals in energy for the two different Ni sites, and presumably resulting in a similarly complicated band structure as in $La_4Ni_3O_{10}$ [24]. As the shape of $NiO_6$ octahedra is altered at the phase transition, the associated changes in the electronic properties must be reflected in a corresponding reconstruction of the Fermi surface. The more pronounced change of the bond lengths within the inner $NiO_6$ octahedra with decreasing temperature as compared to the outer ones may be due to a charge re-distribution between the inner and outer $NiO_6$ octahedra, leading to possible charge and/or orbital order [39, 40] and an insulator-like behavior of the out-of-plane resistivity in the low-temperature phase.

The interpretation that the electronic structure undergoes a substantial change at $T_{pt}$ is supported by the fact that the Curie-constant *C* decreases when crossing $T_{pt}$ from above, and the Pauli-paramagnetic susceptibility as extracted from a Néel-type fit to the magnetic-susceptibility considerably increases [see Table II]. This may indicate that the outer *3d*-electrons of the Ni ions are essentially delocalized at low temperatures, while the increased value of *C* in the high-temperature phase would be compatible with either three localized *S* = ½ magnetic moments, or with one Ni ion in a high-spin *S* = 1 configuration. It is worth noting here that the free-electron value for the Pauli-paramagnetic susceptibility for hypothetically *N* = 3 mobile charge carriers per formula unit (one carrier per nickel atom) corresponds to $\chi_0^{f.e} \approx 10^{-4}$ emu $Oe^{-1}$ $mol^{-1}$, and varies only weakly with *N* as $N^{1/3}$. While this value is compatible with the respective high-temperature values, the significant enhancement below the phase transition points to a substantial change of the electronic structure and a corresponding increase of the electronic density of states at the Fermi level with decreasing temperature.

## V. CONCLUSION

To summarize, we have successfully grown single crystals of $Pr_4Ni_3O_{10}$ with an optical-image floating-zone furnace at high oxygen pressure. We have performed comprehensive experiments in connection with the phase transition at $T_{pt} \approx 157$ K, such as single crystal XRD at different temperatures, heat capacity, electric-transport, and magnetization measurements. While the resistivity data indicate a metal-to-metal transition at $T_{pt}$ within the *a-b* plane and a metal-to-insulator-like transition along the *c*-axis with decreasing temperature, the MR is enhanced right at $T_{pt}$ and exhibits a sign change, which we attribute to a slight suppression of the phase transition by a magnetic field. The first-order nature of this transition and its field dependence $T_{pt}(B)$ can be derived from the combined heat capacity, magnetic-susceptibility and MR data. Both the resistance and magnetic-susceptibility measurements point to a reconstruction of the Fermi surface at $T_{pt}$, with a substantially enhanced Pauli-paramagnetic susceptibility in the more anisotropic low-temperature phase. These changes are likely related to a certain shape variation of the different $NiO_6$ octahedra in the unit cell at $T_{pt}$ upon cooling from high to low temperatures.

## ACKNOWLEDGEMENTS

This work was supported by the Swiss National Science Foundation under Grants No.20-175554, 206021-163997 and PZ00P2-174015, and matching funds from Paul Scherrer Institute for purchasing SCIDRE HKZ (the high-pressure, high-temperature optical-floating zone furnace).

*Supplemental Material for*

# Anisotropic character of the metal-to-metal transition in Pr$_4$Ni$_3$O$_{10}$

Shangxiong Huangfu[*], Gawryluk Dariusz Jakub[†], Xiaofu Zhang[*], Olivier Blacque[‡], Pascal Puphal[†], Ekaterina Pomjakushina[†], Fabian O. von Rohr[‡], Andreas Schilling[*]

[*]Department of Physics, University of Zurich, Winterthurerstrasse 190 CH-8057 Zurich Switzerland

[†]Laboratory for Multiscale Materials Experiments (LMX), Paul Scherrer Institute (PSI), Forschungsstrasse 111, CH-5232 Villigen, Switzerland

[‡]Department of Chemistry, University of Zurich, Winterthurerstrasse 190 CH-8057 Zurich Switzerland

*Synthesis*

Powdered samples of Pr$_4$Ni$_3$O$_{10}$ were synthesized by citric acid assisted sol-gel method. The reactants Pr$_6$O$_{11}$ (99.9%; *Sigma-Aldrich*) and NiO (99.99%; *Sigma-Aldrich*) were weighted in stoichiometric ratios. Then, nitric acid (65% for analysis; *Emsure*) was used for dissolving these oxides, and a clear green liquid was obtained. After adding Citric acid monohydrate (C$_6$H$_8$O$_7$*H$_2$O; 99.5%; *Emsure*) in molar ration 1:1 respect to the molar of cations, the resulting liquid was heated at around 300 ˚C on a heating plate, which was dried and decomposed into a dark brown powder. Finally, an ultrafine and homogeneous powder of Pr$_4$Ni$_3$O$_{10}$ was obtained by annealing the precursor powder at 1100 ˚C in flowing oxygen for 24 hours. Phase purity of the compound was checked with conventional x-ray diffractometer,

*High Oxygen pressure single-crystal growth*

The resulting powder was hydrostatically pressed in the form of rods (5 mm in diameter and ~ 70 mm in length). The rods were subsequently sintered at 1100 ˚C during 24h in oxygen flow.

The crystal growth was carried out using Optical Floating Zone Furnace (HKZ, SciDre, Dresden) with 5000W xenon lamp as a heat source. The growing conditions were the following: the growth rate was 4 mm/h, both rods (feeding and seeding rod) were rotated at about 15 rpm in opposite directions to secure the liquid homogeneity, 140 bar pressure of oxygen was applied during growing.

The oxygen content of the obtained crystal is about 9.96(3) as determined by thermogravimetric reduction with 10% of H$_2$/N$_2$ gas. The single crystal was oriented with Laue Crystal Orientation System (*Photonic Science Inc.*).

*Powder X-ray diffraction*

Powder X-Ray Diffraction (PXRD) data were collected at room temperature in transmission mode using a Stoe Stadi P diffractometer equipped with a CuK$_{\alpha 1}$ radiation (Ge(111) monochromator) and a DECTRIS MYTHEN 1K detector. The reflections of a main phase were indexed with a monoclinic cell in the space group *P2$_1$/a* (No 14). The Rietveld refinement analysis

[1] of the diffraction patterns was performed with the package FULLPROF SUITE [2,3] (version March-2019). The structural model was taken from the single-crystal X-Ray diffraction refinement. Refined parameters were: scale factor, zero shift, transparency, lattice parameters, Pr and Ni atomic positions, and peak shapes as a Thompson–Cox–Hastings pseudo-Voigt function. A preferred orientation correction as a Modified March function was implemented in the analysis.

*Single crystal X-ray diffraction*

The X-ray diffraction data were collected at 293(1) K (room temperature) and 150(1) K on a Rigaku OD XtaLAB Synergy Dualflex (Pilatus 200K detector) diffractometer with an Oxford liquid-nitrogen Cryostream cooler. A single wavelength X-ray source from a micro-focus sealed X-ray tube was used with Mo$K_\alpha$ ($\lambda \approx 0.71073$ Å) radiation and Cu$K_\alpha$ ($\lambda \approx 1.54184$ Å, averaged value of Cu$K_{\alpha 1}$ and Cu$K_{\alpha 2}$) radiation [4]. The selected single crystal was mounted using polybutene oil on a flexible loop fixed on a goniometer head and transferred to the diffractometer. The pre-experiments, data collection, data reduction and analytical absorption correction [5] were performed with the program suite *CrysAlisPro* [6]. Using *Olex2* [7], the structure was solved with the *SHELXT* [8] small molecule structure solution program and refined with the *SHELXL2018/3* program package [8] by full-matrix least-squares minimization on F$^2$. The crystal data and structure refinement parameters are summarized in Table S1, S2 and S3.

*Heat capacity measurements*

The heat capacity of the Pr$_4$Ni$_3$O$_{10}$ single crystal was measured in a Physical Property Measurement System (PPMS, *Quantum Design Inc.*) using a relaxation method in the temperature range between 10 K and 300 K and in external magnetic fields $B = 0$ T and 7 T, for $T$ increasing and decreasing, respectively [see Fig. S1].

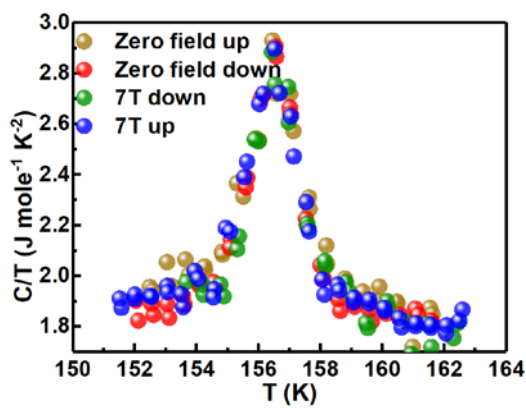

Fig. S1: Reduced heat-capacity $C/T$ as a function of temperature around $T_{pt}$ for $B//c$, measured for $B = 0$ T and 7 T, for increasing ("up") and decreasing ("down") temperatures.

*Transport measurements*

The resistivity measurements of the Pr$_4$Ni$_3$O$_{10}$ single crystals were performed in a PPMS, and a standard four-probe technique was employed with 50 μm silver wires attached with silver paint. All the measurements were performed along both in the *a-b* plane and along the *c*-axis with an applied current of $I = 0.5$ mA in the temperature range from 10 K to 300 K. For the magneto-resistance measurements, the magnetic field was applied perpendicular to the current. As the geometry of the contacts was difficult to control, particularly for the out-of-plane measurements, only estimates of the absolute values could be made. At room temperature, they correspond to roughly $\rho_\parallel \approx 5.6 \times 10^{-3}$ Ωcm and $\rho_\perp \approx 0.047$ Ωcm, respectively.

*Magnetization measurements*

The magnetic properties of $Pr_4Ni_3O_{10}$ were studied with a Magnetic Properties Measurement System (MPMS 3, *Quantum Design Inc.*), equipped with a reciprocating sample option (RSO). We have measured the magnetic moment between 10 K and 300 K, with external fields $B$ parallel and perpendicular to the $c$-axis, and $B$ = 0.01 T, 0.05 T, 0.07 T, 0.1 T, 0.15 T, 0.2 T, 0.5 T, 1 T, 3 T, and 7 T, respectively.

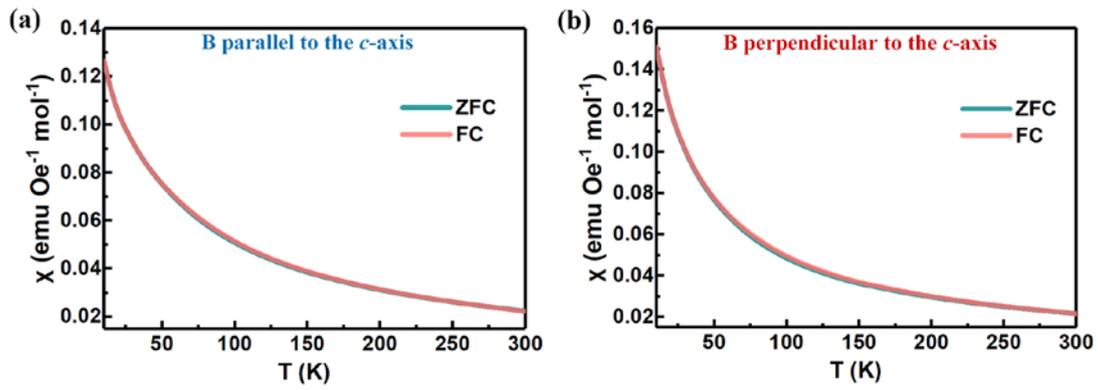

Fig. S2: Zero-field-cooling (ZFC) and field-cooling (FC) magnetization measurement from 10 K to 300 K in an external field of $B$ = 0.1 T; $B$ is (a) parallel and (b) perpendicular to the $c$-axis.

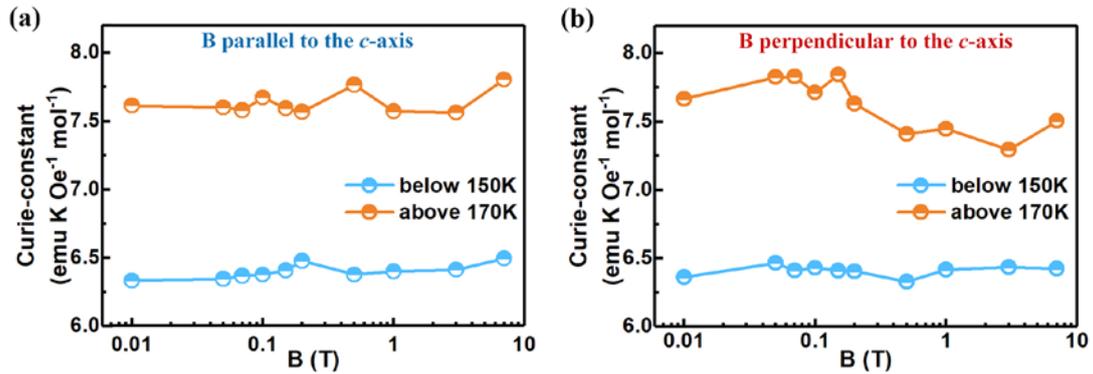

Fig. S3: Curie-constants both above and below $T_{pt}$ for different external magnetic fields, (a) parallel and (b) perpendicular to the $c$-axis, respectively; the corresponding numbers have been obtained from Néel-type fits to the data [see main text].

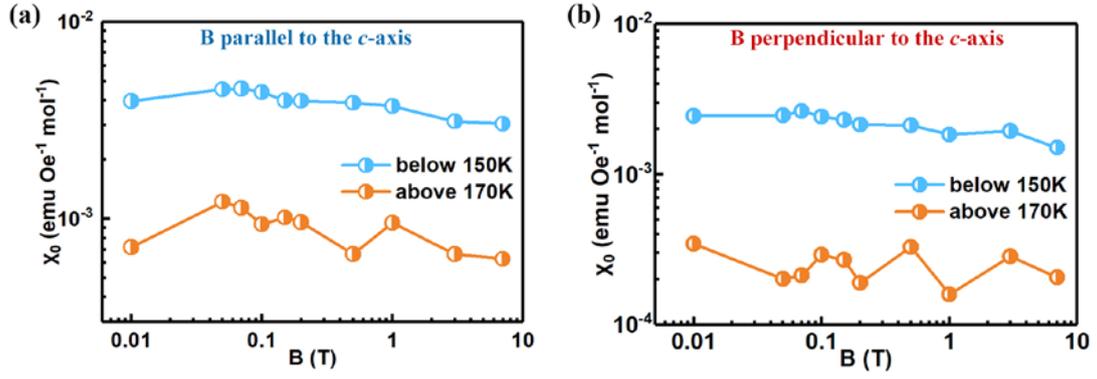

Fig. S4: Pauli paramagnetic-susceptibility from the same fits for temperatures both above and below $T_{pt}$. (a) parallel and (b) perpendicular to the *c*-axis, respectively; the corresponding numbers have been obtained from Néel -type fits to the data [see main text].

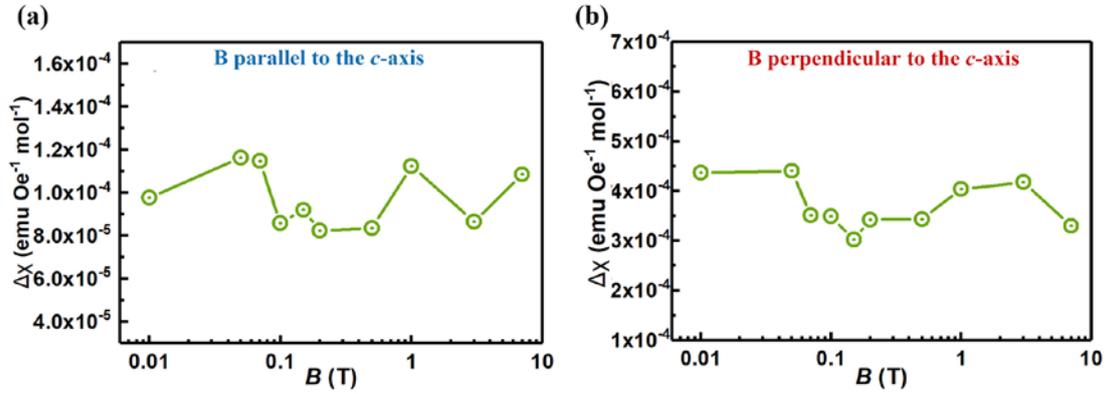

Fig. S5: Step $\Delta\chi$ in the magnetic-susceptibility at the phase transition $T_{pt}$ for different magnetic fields *B*, (a) parallel and (b) perpendicular to the *c*-axis, respectively. [see also Fig. 5b in the main text].

Table SI: Crystallographic data for $Pr_4Ni_3O_{10}$ at different temperatures

| Temperature (K) | 293(1) | 150(1) |
|---|---|---|
| Crystal system | monoclinic | monoclinic |
| Space group | $P2_1/a$ | $P2_1/a$ |
| a (Å) | 5.3771(5) | 5.3664(3) |
| b (Å) | 5.4549(3) | 5.4623(2) |
| c (Å) | 14.028(2) | 13.9889(6) |
| β (°) | 100.910(9) | 100.767(5) |
| V (Å$^3$) | 404.01(6) | 402.83(3) |
| Z | 2 | 2 |
| Radiation type | Mo$K\alpha$, 0.71073 Å | Cu$K\alpha$, 1.54184 Å |
| Absorption coefficient | 30.511 | 189.716 |
| Reflections collected | 5030 | 2691 |
| θ range for data collection (°) | 2.9430 to 31.8700 | 3.1910 to 78.8510 |
| F(000) | 800 | 800 |
| Refinement method | Full-matrix least-squares on F$^2$ | Full-matrix least-squares on F$^2$ |
| Data / restraints / parameters | 1228/18/80 | 724/18/79 |
| Goodness-of-fit | 1.140 | 1.117 |
| $R_1/wR_2$ (I > 2σ) | 0.0874/0.1928 | 0.0716/ 0.2087 |
| $R_1/wR_2$ (all) | 0.0906/0.1936 | 0.0738/ 0.2075 |
| largest diff. peak and hole (Å$^3$) | 8.603 and -11.612 | 3.222 and -4.579 |

Table SII: Atomic positions at 293(1) K

| Atom | Wyckoff site | x | y | z | $U_{eq}$(Å) | occu. |
|---|---|---|---|---|---|---|
| Pr1 | 4e | 0.39828(9) | 0.4889(2) | 0.1983(2) | 0.0055(4) | 1 |
| Pr2 | 4e | 0.1373(2) | 0.4980(3) | 0.0780(3) | 0.0161(4) | 1 |
| Ni1 | 2a | 0 | 0 | 0 | 0.0073(8) | 1 |
| Ni2 | 4e | 0.2789(2) | -0.0022(6) | 0.1406(6) | 0.0060(6) | 1 |
| O1 | 4e | -0.020 (2) | -0.272(3) | 0.221(3) | 0.008(2) | 1 |
| O2 | 4e | 0.296(2) | -0.247(3) | -0.098(4) | 0.014(4) | 1 |
| O3 | 4e | 0.266(2) | 0.249(3) | 0.392(4) | 0.012(3) | 1 |
| O4 | 4e | 0.136(2) | -0.071(3) | 0.069(3) | 0.0064(8) | 1 |
| O5 | 4e | 0.437(2) | 0.037(3) | 0.208(3) | 0.0061(8) | 1 |

Table SIII: Atomic positions at 150(1) K

| Atom | Wyckoff site | x | y | z | $U_{eq}$(Å) | occu. |
|------|--------------|---|---|---|-------------|-------|
| Pr1 | 4e | 0.3981(1) | 0.4872(2) | 0.1985(3) | 0.0109(6) | 1 |
| Pr2 | 4e | 0.1368(1) | 0.4975(2) | 0.0866(3) | 0.0185(6) | 1 |
| Ni1 | 2a | 0 | 0 | 0 | 0.013(2) | 1 |
| Ni2 | 4e | 0.2783(3) | -0.0024(6) | 0.141(1) | 0.013(2) | 1 |
| O1 | 4e | -0.020(2) | -0.272(3) | 0.216(3) | 0.013(2) | 1 |
| O2 | 4e | 0.264(2) | 0.248(3) | 0.402(3) | 0.017(4) | 1 |
| O3 | 4e | 0.297(2) | -0.246(3) | -0.094(3) | 0.020(4) | 1 |
| O4 | 4e | 0.135(2) | -0.066(3) | 0.074(3) | 0.013(2) | 1 |
| O5 | 4e | 0.435(2) | 0.048(3) | 0.208(3) | 0.013(2) | 1 |